\documentclass[reprint,sort&compress,longbibliography,superscriptaddress,amsmath,aps,showpacs,pra]{revtex4-2}
\usepackage{hyperref}
\usepackage{multirow}
\usepackage{amssymb}
\usepackage{amsmath}
\usepackage{graphicx,nicefrac}
\usepackage{color}
\hypersetup{
     colorlinks   = true,
     citecolor    = blue
}
\usepackage{physics}
\usepackage{float}
\usepackage{bbold}
\usepackage[normalem]{ulem}

\graphicspath{{Figs/}}
\newcommand{\be}{\begin{equation}}
\newcommand{\ee}{\end{equation}}
\newcommand{\bea}{\begin{eqnarray}}
\newcommand{\eea}{\end{eqnarray}}

\begin{document}

\title{Designing Moir\'e Patterns by Bending}

\author{Pierre A. Pantale\'on}
\email{panta2mi@gmail.com}
\affiliation{$Imdea\ Nanoscience,\ Faraday\ 9,\ 28015\ Madrid,\ Spain$}
\author{H\'ector Sainz-Cruz}
\affiliation{$Imdea\ Nanoscience,\ Faraday\ 9,\ 28015\ Madrid,\ Spain$}
\author{Francisco Guinea}
\affiliation{$Imdea\ Nanoscience,\ Faraday\ 9,\ 28015\ Madrid,\ Spain$}
\affiliation{$Donostia\ International\  Physics\ Center,\ Paseo\ Manuel\ de\ Lardizabal\ 4,\ 20018\ San\ Sebastian,\ Spain$}

\date{\today}

\begin{abstract}
Motivated by a recent experiment [Kapfer $et$. $al$.,  Science {\bf 381}, 677 (2023)], we analyze the structural effects and low-energy physics of a bent nanoribbon placed on top of graphene, which creates a gradually changing moir\'e  pattern.\ By means of a classical elastic model we derive the strains in the ribbon and we obtain its spectrum with a scaled tight-binding model.\ 
The size of the bent region is determined by the balance between elastic and van der Waals energy, and different regimes are identified.
\ Near the clamped edge, strong strains and small angles leads to one-dimensional channels.\ Near the bent edge, a long region behaves like magic angle twisted bilayer graphene (TBG), showing a sharp peak in the density of states, mostly isolated from the rest of the spectrum.\ We also calculate the band topology along the ribbon and we find that it is stable for large intervals of strains an twist angles.\ Together with the experimental observations, these results show that the bent nanoribbon geometry is ideal for exploring superconductivity and correlated phases in TBG in the very sought-after regime of ultra-low twist angle disorder.
\end{abstract}

\maketitle
\section{Introduction}
Experiments on twisted graphene stacks have unveiled an array of exotic phenomena, including almost all strongly correlated phases known in condensed matter physics~\cite{cao2018correlated,cao2018unconventional,yankowitz2019,lu2019superconductors,polshyn2019large,sharpe2019emergent,serlin2020intrinsic,chen2020tunable,saito2020independent,zondiner2020cascade,wong2020cascade,stepanov2020untying,xu2020correlated,choi2021correlation,rozen2021entropic,cao2021nematicity,stepanov2021competing,oh2021evidence,xie2021fractional,berdyugin2022out,turkel2022orderly,huang2022observation}.\ However, progress in understanding is proving arduous, due to the subtle interplay between factors with comparable energy scales, such as kinetics and Coulomb interactions, angle disorder~\cite{uri2020} and strains~\cite{kazmierczak2021strain}.\ As of now, studying all of these in a single model is out of reach, and experiments struggle with low reproducibility \cite{lau2022reproducibility}.\ Moreover, cascades of correlated phases and superconductivity have been discovered in Bernal bilayer graphene~\cite{Zhou2022SCBG,barrera2022cascade,seiler2022quantum,zhang2022spin,holleis23Ising} and rhombohedral trilayer graphene~\cite{Zhou2021HalfMetRTG,Zhou2021SuperRTG}, crystalline systems with very low strains or disorder, suggesting it may be fruitful to understand these stacks before twisted ones.\ Still, several phases have only been observed in twisted stacks so far, and there is a feeling that the moiré leads to emergent properties. 

Crucial efforts are underway to overcome both angle disorder and strain in twisted graphene systems.\ Strains are commonly found in twisted stacks~\cite{Metal21}, and the combination of twists and strains can lead to many novel moiré structures~\cite{SPG23,Escudero2023Design}.\ In particular, a recent experiment~\cite{Ketal23} has shown that bending 2D materials can create exceptionally uniform moiré patterns, and achieve independent control of twist angle and strain.\ This new technique is promising for next generation experiments on twisted graphene.\ Indeed, samples with ultra-low disorder are expected to enhance already known phenomena, but will also uncover new phases, too delicate to appear in the presence of even mild disorder.\ Reducing disorder from $\sim0.1^{\circ}$ to $\sim0.02^{\circ}$ has allowed for the discovery of zero-field Chern insulators~\cite{stepanov2021competing} and signatures of additional superconducting domes~\cite{lu2019superconductors}, among others.\ Therefore, we are certain that samples with ultra-low disorder $\lesssim0.005^{\circ}$ will be full of surprises.

In this paper, we study the properties of a bent graphene nanoribbon on top of graphene, which creates a slowly changing moiré pattern, as seen in the experiment.\ We first derive the strains in the system using the classical theory of elasticity and then we obtain its spectrum by exact diagonalization of a scaled tight binding model.\ We find that one dimensional channels appear due to the combination of small twist angles and high strains near the clamped edge.\ Moreover, when the bending angle is close to 1$^{\circ}$, a very long region near the bent edge behaves like magic-angle graphene, showing a sharp peak in the density of states, nearly isolated from the rest of the spectrum and an stable band topology.\ These results support the proposal of Ref.~\cite{Ketal23} that bent graphene ribbons are an ideal platform for studying TBG with ultra-low twist angle disorder.\ The rest of the paper is organized as follows: in section II we discuss the elastic properties of the system; \ spectral and topological properties are described in section III, and we conclude in Section IV. 

\begin{figure*}
\includegraphics[scale=0.58]{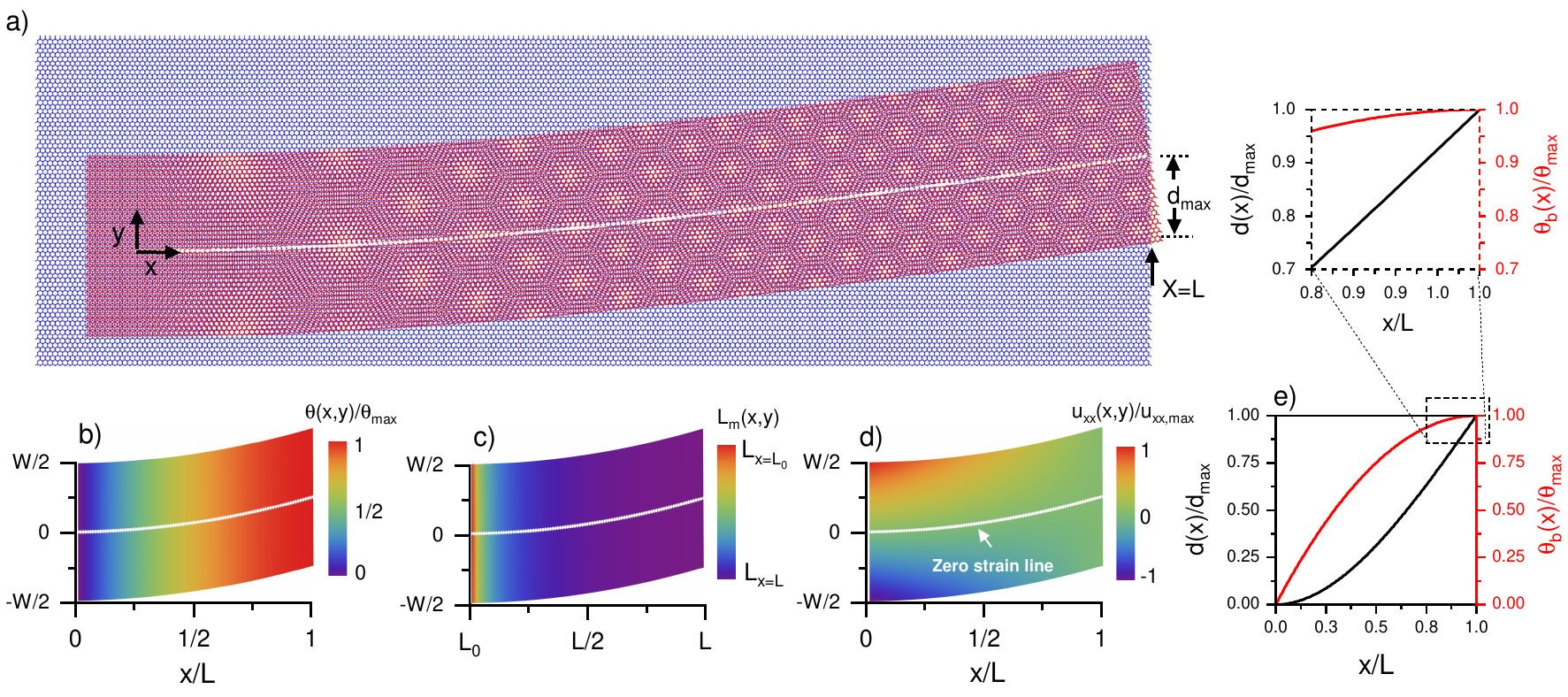}
\caption{(a) Sketch of a bent nanoribbon on top of graphene, which results in a gradually changing moir\'e pattern.\ We highlight the maximum displacement along the y-axes at $x=L$.\ (b) Effective twist angle as function of position, normalized to the maximum twist angle $\theta_{max}$.\ c) Moiré length variation as a function of position, starting from a minimum $L=L_0$ and normalized to its maximum value.\ (d) Strain profile $u_{xx}$ in Eq.~\ref{eq: strainsnano} normalized to the maximum strain value.\ The white line in the middle is the non-strained path.\ e) Nanoribbon deflection curve $d(x)$ and bending angle $\theta_b(x)$ as a function of position, each function is normalized to its maximum value.\ Inset in e) is an enlarged region where a small variation of twist angle is obtained.} 
\label{fig:Figure1}
\end{figure*}

\section{Elasticity}
\subsection{Deformations and stresses}
\label{sec:strains}

We calculate the deformations, $u_x , u_y$ and stresses, $u_{xx}, u_{yy}, u_{xy}$ in a bent nanoribbon of length $L$ and width $W$, in the limit $W \ll L$.\ We assume that one of the narrow sides is clamped  so that:
\begin{align}
u_x ( 0 , y ) &= u_y ( 0 , y ) = 0 , \, \, \, - \frac{W}{2} \le y \le \frac{W}{2}.
\label{clamped}
\end{align}
The ribbon is bent by a vertical force, along the $y$ axis, applied at the other end, $x = L$.\ We adapt to our problem the theory of plates described in~\cite{LL59}, we assume that, to leading order in $W/L$, we can write:
    \begin{align}
        u_y ( x ) &\approx f ( x ) + \cdots
    \end{align}
Then, in the absence of external forces, the elastic energy can be written as:
    \begin{align}
        E_{el} &=  \frac{ \mu ( \lambda + \mu ) W^3}{12  ( \lambda + 2 \mu )} \int_0^L dx \left( \frac{\partial^2 u_y ( x )}{\partial x^2} \right)^2 = \nonumber \\ 
        &= \frac{E W^3}{48} \int_0^L dx \left( \frac{\partial^2 u_y ( x )}{\partial x^2} \right)^2 
        \label{elas0}
    \end{align}
where $\lambda , \mu$ are elastic Lam\'e coefficients, and $E = [ 4 \mu ( \lambda + \mu ) ]/ ( \lambda + 2 \mu )$ is the two dimensional Young modulus.\ In the following, to simplify notation we omit, if necessary, the $x,y$ dependence on the deformation and stress functions.\ The general equilibrium solution of Eq.~\ref{elas} satisfies:
    \begin{align}
        \frac{\partial^4 u_y}{\partial x^4} &= 0.
        \label{der}
    \end{align}
A vertical force, $F$, applied at position $x_0$ of the nanoribbon leads to a term:
    \begin{align}
    \delta E &= F \times \frac{1}{W} \int_{- \frac{W}{2}}^{\frac{W}{2}} u_y ( x_0 , y ) dy
    \label{eq: Fuerza}
    \end{align}
this term induces a discontinuity in $u_y^{iv} ( x ) $ at $x= x_0$.\ In the case of a force applied at the end of the nanoribbon, $x=L$, implies that the function $u_y ( x )$ must have a finite third derivative at $x=L$.\ The clamped condition at $x=0$ also implies that $u'_y ( 0 ) = 0$, so that the most general solution of Eq.~\ref{der} is:
    \begin{align}
        u_y = \frac{a x^2}{2 L} + \frac{b x^3}{3 L^2},
        \label{eq: solutionUy}
    \end{align}
where $a, b$ are dimensionless constants.\ The condition that the shear components of the strain tensor vanish, $\sigma_{x,y} = 0$, in order to set to zero elastic forces at the top and bottom edges of the nanoribbon, implies that:
    \begin{align}
        u_x &= -y \frac{\partial u_y}{\partial x} = - y \left( \frac{a x}{L} + \frac{b x^2}{L^2} \right)
    \label{eq: solutionUx}
    \end{align}
so that:
    \begin{align}
        u_{xx} &= \partial_x u_x = - y \left( \frac{a}{L} + \frac{2 b x}{L^2} \right).  
    \end{align}
The $\sigma_{yy}$ component of the stress tensor should also vanish, and:
    \begin{align}
    \sigma_{yy} &= \lambda ( u_{xx} + u_{yy} ) + 2 \mu u_{yy} = 0,
        \nonumber \\
        u_{yy} &= - \frac{\lambda}{\lambda + 2 \mu} u_{xx} = - \nu u_{xx},
    \end{align}
where $\nu = \lambda / ( \lambda + 2 \mu )$ is the Poisson ratio.\ Equation~\ref{eq: solutionUy} requires a correction of order $W/L$:
    \begin{align}
        \delta u_y &= - \frac{\nu y^2}{2}  \left( \frac{a}{L} + \frac{2 b x}{L^2} \right).
    \end{align}
The only non zero component of the stress tensor is:
    \begin{align}
        \sigma_{xx} &= \lambda ( u_{xx} + u_{yy} ) + 2 \mu u_{xx} =
        \nonumber \\ 
        &=
        \frac{4 \mu ( \lambda + \mu )}{\lambda + 2 \mu} y \left( \frac{a}{L} + \frac{2 b x}{L^2} \right),
    \end{align}
a quantity that must vanish at the non clamped edge of the nanoribbon, $x=L$, so that:
    \begin{align}
        b &= - \frac{a}{2}.
    \label{eq: relaBA}    
    \end{align}
The angle between the bent nanoribbon and the horizontal axes is given by 
\begin{align}
        \theta(x,y) = \frac{\partial u_y (x,y)}{\partial x}.
        \label{eq: angulo}
    \end{align}
The maximum twist angle occurs for $x=L$ and it is:
    \begin{align}
        \theta_{max} &\approx \left. \frac{\partial u_y (x,y)}{\partial x} \right|_{x=L}\approx a + b = \frac{a}{2},
    \end{align}
so that $a = -2 b \approx 2 \theta_{max}$.\ The strains inside the nanoribbon are:
    \begin{align}
        u_{xx} (x,y) &= - \theta_{max} \frac{y}{L} \left( 1 - \frac{x}{L} \right), \nonumber \\
        u_{yy} (x,y) &= - \nu u_{xx} (x,y).
        \label{eq: strainsnano}
    \end{align}
Note that we have replaced the force applied at the free end by the value of the resulting twist angle. The deformations satisfying the boundary conditions are then given by
 
\begin{align}
    u_x (x,y) &= -2 \theta_{max}\frac{y}{L}\left(x - \frac{x^2}{2 L}\right), 
    \label{eq: Ux} \\
    u_y (x,y) &= \frac{\theta_{max}}{L} \left[\left(x^2 - \frac{x^3}{3 L}\right) - \nu y^2 \left(1 - \frac{x}{L}\right)\right],
    \label{eq: Uy}
\end{align}
which are the local displacements of each point within the nanoribbon. From Eq.~\ref{eq: angulo} and Eq.~\ref{eq: Uy} we obtain an explicit form for the  twist angle,  
\begin{equation}
    \theta(x,y)=\frac{\theta_{max}}{L}\left(2x-\frac{x^{2}}{L}-\frac{y^{2}\nu}{L^2}\right), 
\label{eq: anglexy}
\end{equation}
between the nanoribbon and the horizontal axis. 

\begin{figure}[t]
\includegraphics[scale = 0.75]{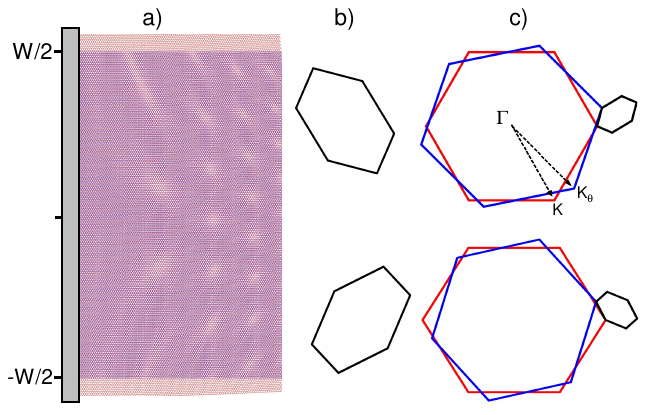}
\caption{(a) Moir\'e pattern of the bent nanoribbon in a region close to the clamped edge. (b) Real space unit cell for a given value of strain and twist angle.\ (c) Formation of the mini Brillouin zone (mBZ, black) due to the uniform substrate (red) and the twisted and strained lattice (blue).\ The orientation of both real unit cell and mBZ depends on the combination of twist angle and strain sign and magnitude.
}
\label{fig:Figure2}
\end{figure}

\subsection{Clamping and sliding.}
The previous subsection describes the deformations, strains, and stresses of a nanoribbon with one of the ends clamped and the other under a lateral force. We now estimate, as function of the van der Waals forces between the nanoribbon and the substrate, which fraction of the nanoribbon remains clamped to the substrate, and which fraction slides because of the force applied to the end. The size of the bent region is determined by a balance between the interlayer van der Waals coupling, and the elastic energy required to bend the nanoribbon. 

The elastic energy of the bent nanoribbon can be easily calculated from the estimates of the strains  discussed previously. In the limit $W \ll L$, the elastic energy scales as:
\begin{align}
    E_{elas} &\propto \mu W^3 \int_0^L \left( \frac{\partial^2 u_y}{\partial x^2} \right)^2 d x \sim  \frac{\mu W^3 \theta_{max}^2}{L}.
    \label{elas}
\end{align}

The van der Waals interaction can be described, in a first approximation, by the energy per unit area between of perfectly aligned layer, $v_{vdW}$, and the assumption that, in the limit of large misalignment, the two layers are decoupled.

The van der Waals energy between the nanoribbon and the substrate can be approximated by the energy difference, per unit area, between perfect alignment between the two layers, and the energetically less favorable alignment, denoted as $V_{vdW}$. These two configurations repeat themselves with a periodicity comparable to the intralayer distance between nearest neighbor atoms. An interpolation using a few harmonics gives a rough description of the van der Waals interaction. The value of $V_{vdW}$ is of order 1 meV\AA$^{-2}$.

In the following, we estimate the van der Waals energy for two possible regimes:

\subsubsection{Complete decoupling between the nanoribbon and the substrate.}
We assume perfect alignment of the clamped region with the substrate, with the bent region considered decoupled from it. The van der Waals energy required to detach the bent region is 
\begin{align}
E_{vdW} ( L ) &= V_{vdW} \times ( L W ).
\label{vdW1}
\end{align}
The total energy needed to bend the nanoribbon is determined by the sum of this term and the elastic energy, Eq.~(\ref{elas}). We approximate $\theta_{max} \approx d_{max} / L$, where $d_{max}$ is the vertical deflection at the edge. Then, the optimal value of $L$, the length of the detached region is given by:
\begin{align}
    L &\approx \left( d_{max} W \right)^{1/2} \left( \frac{\mu}{V_{vdW}} \right)^{1/4}.
\end{align}

\subsubsection{Misaligned substrate and partial relaxation.
}

We consider that neither the clamped nor the bent regions are fully aligned or misaligned with the substrate. We assume that the misalignment is described by a twist angle, $\theta_{subs}$ in the clamped region, which changes smoothly in the bent region: 
\begin{align}
\theta ( x ) & \approx \theta_{subs} + \theta_{max} \left( \frac{x}{L} \right)^2.
\end{align}
At each position, $x$, in the nanoribbon a moiré pattern can be defined, associated to the twist angle $\theta ( x )$. The van der Waals force between the nanoribbon and the substrate leads to a relaxation, and to  a finite attractive van der Waals interaction. The force per unit area exerted by the substrate on the nanoribbon is of order $V_{vdW} / \ell$, where $\ell$ is comparable to the interatomic distance in each layer. The deformation induced by this force arises from the balance of this force and the cost in elastic energy associated to the relaxation of the atomic positions. The moiré defined by the twist angle $\theta ( x )$ is of order $\ell_m ( x ) \sim \ell / \theta ( x )$. The wavelength of 
 the induced strains is of order $\ell_m ( x  )$, which leads to an effective spring constant, per unit area, of order $\mu / \ell_m ( x )^2$. Using second order perturbation theory, the attractive van der Waals energy, per unit area is 
 \begin{align}
\tilde{E}_{vdW} ( x ) &\propto  - \frac{V_{vdW}^2}{\mu} \frac{\ell_m ( x )^2}{\ell^2} \approx   \frac{V_{vdW}^2}{\mu \theta ( x )^2}   
 \end{align}
 The total van der Waals energy can be obtained by integrating this expression over $x$, from $x=0$ to $x=L$. 
 
We now assume that the initial misalignment in the clamped region is much larger than the change induced by bending, $\theta_{max} \ll \theta_{subs}$. Then, we can expand the van der Waals energy as:
 \begin{align}
 E_{vdW} &= \int_0^L \tilde{E}_{vdW} ( x ) dx \approx \nonumber \\
 &\approx - \frac{V_{vdW}^2 W}{\mu} \left( c_1 \frac{L}{\theta_0^2} + c_2 \frac{d_{max}}{\theta_0^3} + c_3 \frac{d_{max}^2}{L \theta_0^4} + \cdots \right).
 \label{vdW2}
\end{align}
where $c_1 , c_2 , c_3$ are dimensionless constants of order unity, and we are using $\theta_{max} \approx d_{max} / X$. The size of the bent region arises from the balance of this expression and the elastic energy, Eq.~\ref{elas}. The first term in Eq.~\ref{vdW2} is independent of the bending, and the second term does not contain the length $L$. From the third term in Eq.~\ref{vdW2} and Eq.~\ref{elas}, we find:
\begin{align}
    L &\approx W \theta_0^2 \, \frac{\mu}{V_{vdW}}. 
    \label{bent}
\end{align}
This regime seems to approximately describe the experiments reported in~\cite{Ketal23}, as it was found that $L \propto W$, and the values of $L$ and $d_{max}$ seem to be independent. The ratio $L / W \approx 7$ reported in~\cite{Ketal23} is consistent with $\mu / V_{vdW} \approx 10^4$ and $\theta_0 \approx 1.5^\circ$. Note, however, that Eq.~\ref{bent} suggests a significant dependence on the misalignment of the clamped region, parametrized by $\theta_0$, which could imply a sample dependence not observed in the experiments.

\subsection{Bent graphene nanoribbons}

The model presented in the previous sections is applicable for determining the elastic properties of nanoribbons with an arbitrary geometry.\ However, recent experiments have realized bent graphene nanoribbons placed on a graphene substrate~\cite{Ketal23}.\ A sketch of the system is shown in Fig.~\ref{fig:Figure1}(a).\ The position of each lattice site is given by the set $\{x_i,y_i\}$ for graphene and $\{x_i+ u_x (x_i,y_i),y_i+ u_y (x_i,y_i)\}$ for the bent nanoribbon, where index $i$ runs over all positions in the honeycomb lattice.\ In the following, without loss of generality, we are assuming an initial non twisted $AB$ stacking at the clamped region (see Fig.~S1).\ Near this edge, the combination of high strains and low angles leads to a strong deformation of the moiré lattice and the formation of quasi-one-dimensional patterns~\cite{SPG23,Escudero2023Design}.\ A zoom near this zone is shown in Fig.~\ref{fig:Figure2}(a), where the bright spots correspond to $AA$ regions.\ The different sign of the strains inside the nanoribbon, Eq.~\ref{eq: strainsnano}, results in a different distortion of both real and reciprocal space, as shown in Fig.~\ref{fig:Figure2}(b, c), respectively.\ The sign and strength of the strain can be inferred from the distortion of the $AA$ regions~\cite{Escudero2023Design}. 

For regions far from the clamped edge, the deformations in Eq.~\ref{eq: Ux} and Eq.~\ref{eq: Uy} give rise to a smooth sequence of moir\'e patterns whose dimensions become almost uniform.\ Figure~\ref{fig:Figure1}(b) shows the variation of the twist angle as a function of position.\ Close to the clamped edge, the angle grows linearly while it is nearly uniform close to the bent edge, cf.\ inset in Fig.~\ref{fig:Figure1}(e).\ Notice that there is also a small transversal variation of the twist angle, cf.~Eq.~\ref{eq: anglexy}.\ For large nanoribbons the variation of the twist angle becomes minimal as we move away from the clamped edge.\ This angle uniformity can also be observed in the dependence of the moiré length with position, which can be obtained with the local twist angle and the components of the strain tensor in Eq.~\ref{eq: strainsnano}.\ Figure~\ref{fig:Figure1}c) shows the variation of the moire length with position.\ The larger values are close to the clamped edge where the combination of small twist and strains gives rise to a complicated lattice structure, shown in Fig.~\ref{fig:Figure2}. As we move away from the clamped edge the moire length becomes almost perfectly uniform.\ Figure~\ref{fig:Figure1}d) displays the spatial profile of the $u_{xx}(x,y)$ component of the strain tensor, cf. Eq.~\ref{eq: strainsnano}.\ This component changes linearly in both the longitudinal and transverse directions.\ The highest strain occurs near the edges of the clamped side.\ The magnitude of the strain\ components is symmetric around the zero strain line, and its sign is determined by the direction of the applied force.\ This leads to compression (expansion) on the upper (lower) side of the nanoribbon.

On the other hand, the zero strain line, shown in white in Fig.~\ref{fig:Figure1}, is given by the positions $\{x_i,u_y (x_i,0)\}$, where the second component is a displacement in the $y$-direction and is called {\it deflection curve}, $d(x)$, which is a function describing the nanoribbon bending along the vertical direction~\cite{LL59}.\ The normalized deflection curve $d(x)/d(L)$ is universal and describes the shape of any bent nanoribbon.\ A plot of this function is shown in Fig.~\ref{fig:Figure1}(e).\ The variation of the deflection curve in a given position determines the bending angle $\theta_b(x)= \theta(x,0)$ which is the twist angle, with respect to the graphene substrate, along the zero strain line.\ In an experimental setup, for a nanoribbon of length $L$, the deflection curve gives the bending angle which determines the maximum twist angle $\theta_{max}$.\ By knowing the length and the maximum twist angle (or the maximum of the deflection curve), thus Eq.~\ref{eq: Ux} and Eq.~\ref{eq: Uy} determine the full elastic properties.\ In addition, the enlarged plot in Fig.~\ref{fig:Figure1}(e) shows a region where the twist angle is almost uniform.\ Indeed, in this region, the bent nanoribbon has been shown to have very low disorder, with small variations of twist angles and strain~\cite{Ketal23}.

\subsection{Numerical estimates.}

As previously described, the strains in the bent nanoribbon are highest at the edges, $y = \pm W/2$ and they are antisymmetric around the center, $y = 0$.\ The maximum values occur at $x=0$,
\begin{align}
 u_{xx} \left( 0, \pm \frac{W}{2} \right) &= \pm \theta_{max} \frac{W}{L}, \nonumber \\
 u_{yy} \left( 0, \pm \frac{W}{2} \right) &= \mp \nu \theta_{max} \frac{W}{L}. 
\label{eq:MaxStrains}
\end{align}
It is interesting to note that the moiré pattern defined by a combination of a twist and an uniaxial strain leads to quasi one dimensional behavior~\cite{SPG23, Escudero2023Design} when $- u_{xx} u_{yy} \approx \theta^2$.\ Using the expression presented above, this relation is satisfied for 
    \begin{align}
    \frac{y}{L} &= \frac{\frac{x}{L} \left( 1 - \frac{x}{2 L} \right)}{\sqrt{\nu} \left( 1 - \frac{x \sqrt{\nu}}{L} \right)} \approx \frac{x}{L \sqrt{\nu}},
     \end{align}
and the geometrical effects can be observed in the distortions in  Fig.~\ref{fig:Figure2}.  

In realistic graphene nanoribbons on a graphene substrate~\cite{Ketal23}, the quotient $W/L \sim 0.10$ and the maximum twist angle $\theta_{max}\sim 2.5^\circ$ results in a strain tensor, Eq.~\ref{eq:MaxStrains}, with components of magnitude $u_{xx} = \pm 0.43 \%$  and  $u_{yy} = \pm 0.72\%$ for a Poisson ratio of $\nu =  0.165$.\ These values are in excellent agreement with the mapping of the strain profile in Ref.~\cite{Ketal23}.\ On the other hand, both the deflection curve and the bending angle tend to be uniform.\ The uniform twist angle region is that where $x/L \gtrsim 0.8$, as shown in Fig.~\ref{fig:Figure1}. 

Before introducing the electronic structure, it is important to note that the analyzed strains induce an effective gauge field that interacts with the electrons.\ We assume that the lattice axes are oriented such that the gauge field can be written as (note that the sign depends on the valley):
\begin{align}
    \{ A_x , A_y \} &= \pm \frac{3\beta}{2a} \times \left\{ \begin{array}{lr} \{ -u_{xx} + u_{yy} , 2 u_{xy}  \}
    & {\rm zigzag}\\
    \{ 2 u_{xy} , u_{xx} - u_{yy}  \} & {\rm armchair}
    \end{array} \right.
\end{align}
where $\beta = (a/t) ( \partial t / \partial a )$ is a dimensionless constant that describes the change of the nearest neighbor hopping parameter, $t$, with respect to the inter atomic distance, $a$.\ In the present case, we obtain:
\begin{align}
    \{ A_x , A_y \} &= \pm \frac{3\beta}{2a} ( 1 + \nu ) \times \left\{ \begin{array}{lr}  \{ - u_{xx} , 0   \} & {\rm zigzag} \\
   \{  0 , u_{xx} \} & {\rm armchair}
  \end{array} \right.
\end{align}
The effective magnetic field is:
\begin{align}
    B ( x , y ) &= \frac{1}{\ell_m^2} = \frac{\partial A_y}{\partial x} - \frac{\partial A_x}{\partial y} =
    \nonumber \\ 
    &= \pm \frac{3\beta}{2a} ( 1 + \nu ) ( - 2 \theta_{max} ) \times \left\{  \begin{array}{lr}   - \frac{1}{L} + \frac{x}{L^2} & {\rm zigzag}\\
 \frac{y}{L^2} 
 & {\rm armchair}
    \end{array} \right.
\end{align}
where $\ell_m$ is the magnetic length.
The highest possible magnetic field occurs for the zigzag orientation at the clamped edge.\ Then, the magnetic length is:
\begin{align}
    \ell_m &\approx   \sqrt{\frac{L a}{3 \theta_{max} \beta ( 1 + \nu )}}
\end{align}
For $\theta_{max} \sim 4^\circ $ and $L \sim 10 \mu$m we obtain $\ell_m  \sim 500$nm.\ The corresponding magnetic fields are below 1T, so that the effect on the electrons is  small.

\section{Spectrum and topology}
\subsection{Spectrum}

We now calculate the electronic spectrum of the system, using a tight-binding Hamiltonian~\cite{lin2018minimum} with a scaling approximation~\cite{gonzalezarraga2017electrically,vahedi2021magnetism,sainzcruz2021high}, for details see Ref.~\cite{SM}. We compute the spectrum of a bent nanoribbon like the one shown in Fig.~\ref{fig:Figure1}(a), but with a length of $L\approx800$ nm and a width of $W\approx$ 80nm, after scaling.\ There are $N\approx2.6\cdot10^5$ sites.\ The geometric twist angle changes from 4.4$^{\circ}$ to 0$^{\circ}$.\ However, thanks to the scaling approximation, we can use this lattice to simulate a change from 1.1$^{\circ}$ (near the magic angle) to 0$^{\circ}$ (cf. SM Sec.~\ref{App: Ribbon}).\ Near the bent edge, the system has twist angles near the magic angle and little strain, while at the clamped edge there is an interplay between small angles and large strains, which leads to one-dimensional channels, that are revealed by the lattice structure in Fig.~\ref{fig:Figure2}(a) and the charge map in Fig.~\ref{fig: Figure6} below.

Figure~\ref{fig: Figure4} shows the DOS of the system.\ We compare the DOS across various regions: the yellow line corresponds to states near the clamped edge on the left side ($\theta\approx0^{\circ}-0.57^{\circ}$), the red line corresponds to states near the right end ($\theta\approx 0.98^{\circ}-1.10^{\circ}$), the black line represents the total DOS across the entire ribbon and in the blue line we consider the contribution of the monolayer fringes (cf. SM Sec.~\ref{App: Ribbon}).\ The low-energy DOS near the bent edge (red curve) has a prominent peak while the DOS near the left end (yellow curve) is negligible in comparison.\ The peak is due to states localized at the AA stacking, see e.g. the charge map in Fig.~\ref{fig: Figure6}, and it is not at zero energy due to electron-hole asymmetry and to a rigid blue-shift induced by scaling. We observe that a long region starting at the bent edge of the nanoribbon behaves like low-disorder magic-angle TBG, as evidenced by the sharp peak in the density of states.\ Moreover, on both sides of the peak, the DOS is almost zero, hinting at the presence of gaps to the rest of the spectrum, as it happens when flat bands are separated from remote bands.\ Therefore, correlated phases and superconductivity, in as much as they depend on these features, should also be present.\ Moreover, the system studied here is much smaller than the experimental one, on which the idea works even better, thanks to the many unit cells and very slow changing twist angle. 

\begin{figure}
\includegraphics[scale=0.55]{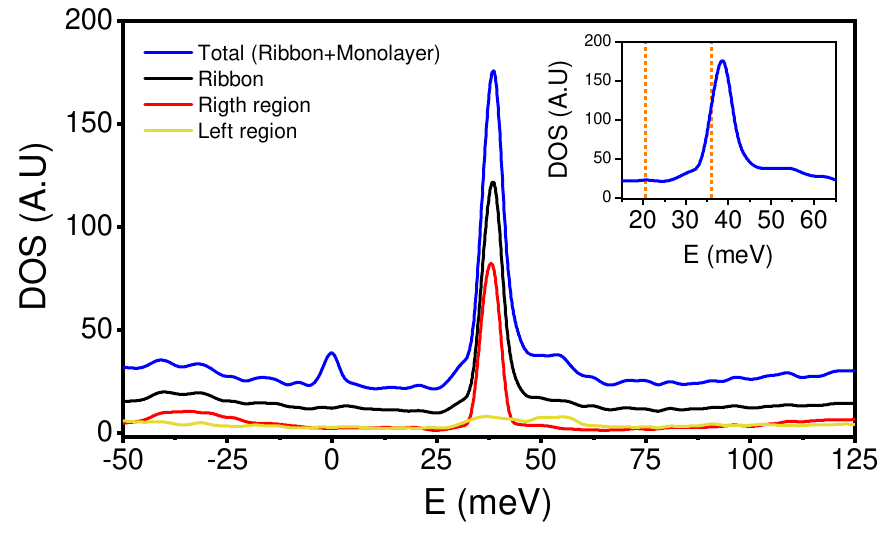}
\par
\caption{Low energy density of states (DOS) of a system like the one in Fig.~\ref{fig:Figure1}, including the monolayer fringes (blue), versus the DOS of the ribbon (black) and the local DOS at the right end of the ribbon, spanning angles [$0.98^{\circ}$, $1.1^{\circ}$] (red) and at the left end ([$0^{\circ}$, $0.57^{\circ}$], yellow). Inset is the total DOS indicating the energy positions (dashed lines) of the charge density maps on Figs.~\ref{fig: Figure6} and~\ref{fig: Figure7}.  
}
\label{fig: Figure4}
\end{figure}

As a side note, the middle region, which has twist angles in the range $\theta\approx0.4^{\circ}-0.7^{\circ}$, also shows localization near the AA stacking regions~\cite{fleischmann2018moire}, with {\it annular} shapes instead of peaks for lower angles, see Fig.~\ref{fig: FigureS10}.\ We note that this annular behavior of the charge density closely resembles the charge localization for different magic angles as described in Refs.~\cite{Navarro2022WhyMagic,Navarro2023Quantization}.\ Furthermore, the small peak near zero energy in Fig.~\ref{fig: Figure4} originates from the edge states that are a consequence of the finite boundary conditions of the monolayer fringes~\cite{Wakabayashi2010}. These edge states can also manifest at twisted bilayer boundaries~\cite{Yin2022Direct,Andrade2023}.

The results in Fig.~\ref{fig: Figure4} are a clear indication that the bent nanoribbon geometry pioneered in Ref.~\cite{Ketal23} is useful for observing magic angle TBG physics and searching for new phenomena in the uncharted regime of with ultra-low angle disorder.\ For optimal results, the bending angle should be close to the magic angle.\ This configuration will benefit from the slow changing angle near the bent edge, combined with low strains.

\begin{figure*}
\includegraphics[scale = 0.85]{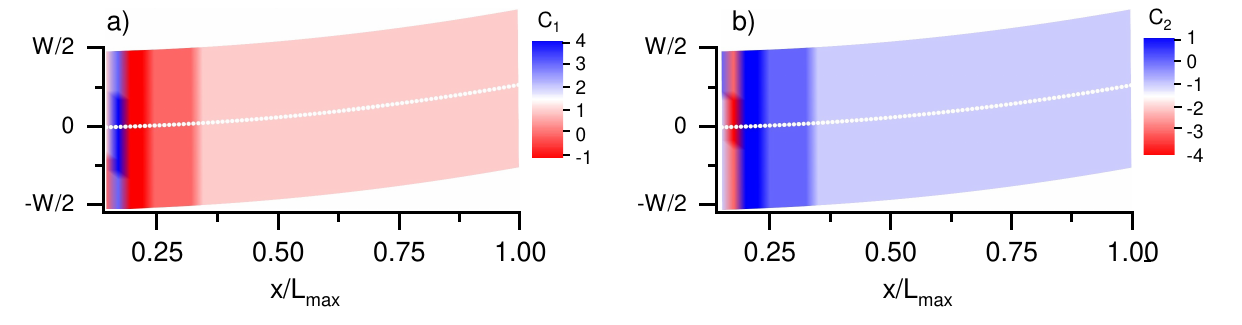}
\caption{Topological phase diagram of a bent graphene nanoribbon.\ The maximum twist angle is set to $\theta_{max} = 1.5^\circ$ with Poisson ratio $\nu = 0.165$.\ Topological phases for the a) lower and b) upper middle bands.\ The corresponding Chern numbers are shown in colors. 
}
\label{fig:topofase}
\end{figure*}

\subsection{Topology}
\begin{figure*}
\includegraphics[scale=0.7]{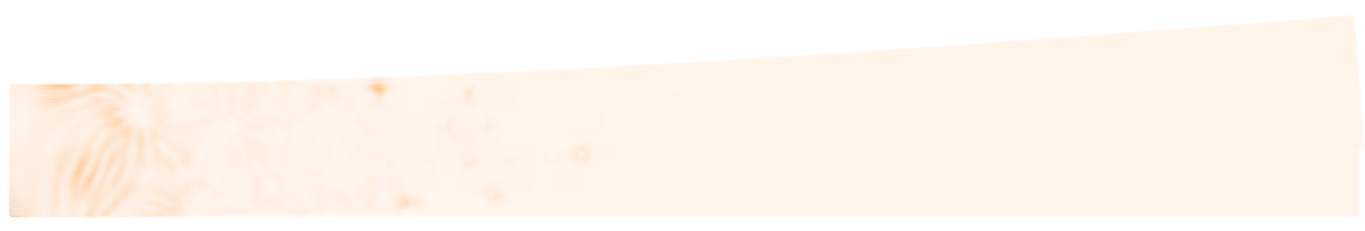}
\par
\caption{Charge map of a state with $E=20.58$ meV. The panel on the top is the entire system. On the left side, the combination of high strains and twists gives rise to charge stripes.}
\label{fig: Figure6}
\end{figure*}

In a large nanoribbon, the slow variation of twist angle and strain allows us to locally calculate its electronic properties.\ As described in the previous section, the narrow bands dominate giving rise to a large density of states.\ 
In experiments, the TBG samples are supported on~\citep{Serlin2019} or encapsulated~\cite{Setal20,Ma2020Enc} in hexagonal boron nitride (hBN).\ In the following, we consider the presence of a mass term in the bent graphene nanoribbon.\ This can be achieved, for example, by considering an hBN substrate acting only on the bottom graphene layer~\cite{Misc3}.\ The hBN substrate induce a mass term that breaks inversion symmetry allowing for a finite Berry curvature~\cite{Jung2015}.\ We focus on a large nanoribbon, enabling the local determination of its electronic characteristics.\ Band topology is evaluated over a range of bending (or twist) angles, varying from around $\theta\approx 0.2^\circ$ to $\theta\approx 1.5^\circ$.\ We obtain the valley Chern numbers of the two low-energy central bands as a function of the local twist angle, cf. Eq.~\ref{eq: anglexy} and local strain tensor, cf. Eq.~\ref{eq: strainsnano}, using a continuum model of strained twisted bilayer~\cite{SPG23,Pantalen2022}.\ We also introduce a mass term of $\Delta = 15$ meV~\cite{Long2022}.\ As shown in Fig.~\ref{fig:topofase}(a) and Fig.~\ref{fig:topofase}(b), near the clamped edge there are different topological phases with valley Chern numbers varying from $\mathcal{C} = \pm 4$ to $\mathcal{C} =\pm 1$.\ This result highlights the intricate nature of the bands when both low twist and strain are present.\ This complex behavior is also reflected in the charge map displayed in Fig.~\ref{fig: Figure6} and the geometric profile depicted in Fig.~\ref{fig:Figure2}.\ It is important to underscore that the electronic structure near the clamped side exhibits high sensitivity to the chosen parameters. A comprehensive study may be necessary to thoroughly characterize the complete electronic properties in this region.\ Moreover, within the nanoribbon, certain regions exhibit clearly defined topological transitions that display reduced sensitivity to variations in parameters.\ On the left side of the ribbon, specifically concerning the two middle bands, the phase characterized by Chern numbers $\{ \mathcal{C}_1, \mathcal{C}_2 \} = \{-1,1\}$ is found within the range of $\theta \sim \left(0.49^{\circ},0.63^{\circ}\right)$ and $x/L \sim \left(0.18,0.24\right)$.\ The phase with $\{ \mathcal{C}_1, \mathcal{C}_2 \} = \{0,0\}$ occurs within $\theta \sim \left(0.63^{\circ},0.86^{\circ}\right)$ and $x/L \sim \left(0.24,0.35\right)$.\ The dominating phase from the central to the right end is the one characterized by Chern numbers $\{ \mathcal{C}_1, \mathcal{C}_2 \} = \{1,-1\}$, existing for $\theta \gtrsim 0.86^{\circ}$ and $x/L \gtrsim 0.35$.\ This is the well known topological phase of TBG on an hBN substrate~\cite{Long2022,Shi2021Conm, Cea2020TBGhbN, Shin2021asy, Mao2021Quasi}.\ Our findings are consistent with the smooth variation of twist angle and moir\'e length described in the previous sections, cf.\ Fig.~\ref{fig:Figure1}.\ Thus, in an experimental setup, our results indicate that the uniform strains and small variations of twist angles in large regions, give a uniform electronic structure and band topology, further supporting the proposal of Ref.~\cite{Ketal23} that bent graphene nanoribbons on a graphene substrate are ideal platforms to study TBG with ultra-low angle disorder.  

\begin{figure*}
\includegraphics[scale=0.7]{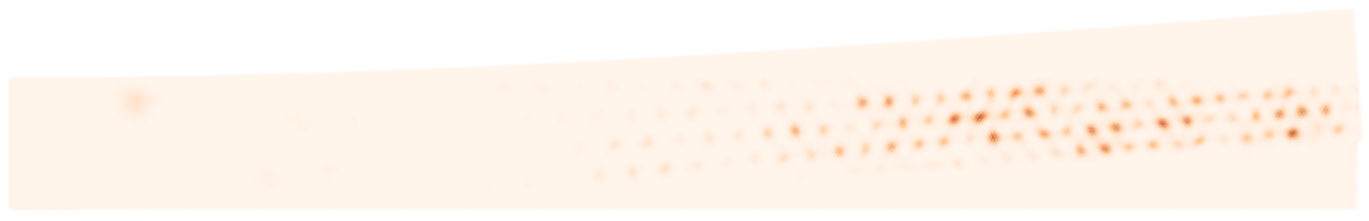}
\par
\caption{Charge map of a state with $E=38.99$ meV, showing charge localization at most AA regions near the right edge. States like this one make the peak in the DOS seen in Figure \ref{fig: Figure4}. 
}
\label{fig: Figure7}
\end{figure*}

\section{Conclusions}

In this paper, we have examined the elastic and spectral characteristics of a graphene nanoribbon that has been bent on top of graphene.\ This arrangement has been recently achieved experimentally~\cite{Ketal23}.\ This geometry offers a crucial advantage by allowing independent control of twist angle and strain while reducing angle disorder.\ By means of the classical theory of elasticity, we have analyzed the deformations and stresses within the system, which result in a gradual change in the twist angle and the corresponding moiré pattern. 

The size of the bent region is determined by a balance between the elastic energy, which for fixed shift of the free end decreases with length, and the reduction in van der Walls energy, which increases with the length. Different regimes have been identified.

\ In proximity to the clamped edge, the combination of strong strains and small angles induces quasi-one-dimensional channels, aligning with earlier findings~\cite{SPG23}.\ As a consequence, states in these region can show charge stripes. 


In a long region at the right side of the bent edge, strains and twist angles are nearly uniform, as seen in the experiment.\ In this region, the low-energy spectrum includes many states with charge localization in the AA stacking regions which lead to a sharp, nearly isolated, peak in the density of states, as expected for magic angle twisted bilayer graphene.\ We attribute this to the existence of flat bands in a wide range of twist angles and strains.\ We have also calculated the band topology and found that the Chern number depends only on the twist angle for large intervals of strains and angles.\ Our results suggest that a bent nanoribbon geometry pioneered in Ref.~\cite{Ketal23} is ideal for exploring superconductivity and correlated phases in TBG in the very sought-after regime of ultra-low angle disorder.
\vfill\null

\subsection*{Acknowledgements}

We thank Rebeca Ribeiro-Palau, M. Kapfer, B. Jessen, Federico Escudero, and Zhen Zhan for discussions.\ IMDEA Nanociencia acknowledges support from the \textquotedblleft Severo Ochoa\textquotedblright~Programme for Centres of Excellence in R\&D (CEX2020-001039-S/AEI/10.13039/501100011033).\  We acknowledge funding from the European Commission, within the Graphene Flagship, Core 3, grant number 881603 and from grants NMAT2D (Comunidad de Madrid, Spain), SprQuMat, (MAD2D-CM)-MRR MATERIALES AVANZADOS-IMDEA-NC, NOVMOMAT, Grant PID2022-142162NB-I00 funded by MCIN/AEI/ 10.13039/501100011033 and by “ERDF A way of making Europe”.

\bibliography{references}

\begin{thebibliography}{62}
\bibitem [{Mis()}]{Misc3}%
  \BibitemOpen
  \href@noop {} {}\bibinfo {note} {An hBN substrate can also induce a moiré
  pattern in the bottom graphene monolayer which may result in a moir\'e
  pattern that can affect the band structure due to an effective trilayer
  structure. This effects requires a detailed investigation that we leave for a
  future discussion.}\BibitemShut {Stop}%
\end{thebibliography}
\bibliographystyle{apsrev4-1}
\clearpage

\setcounter{equation}{0}
\setcounter{figure}{0}
\setcounter{table}{0}
\setcounter{page}{1}
\setcounter{section}{0}
\makeatletter
\renewcommand{\theequation}{S\arabic{equation}}
\renewcommand{\thefigure}{S\arabic{figure}}

\onecolumngrid

\vspace{1cm}
\begin{center}

{\Large Supplementary information \textit{for} \\ Designing Moiré Patterns By Bending} \\
\vspace{0.25cm}
Pierre A. Pantale\'on, H\'ector Sainz-Cruz and Francisco Guinea
\end{center}

\section{Bent nanorribbon} \label{App: Ribbon}
\begin{figure*}[h]
\centering
{\includegraphics[scale = 0.59]{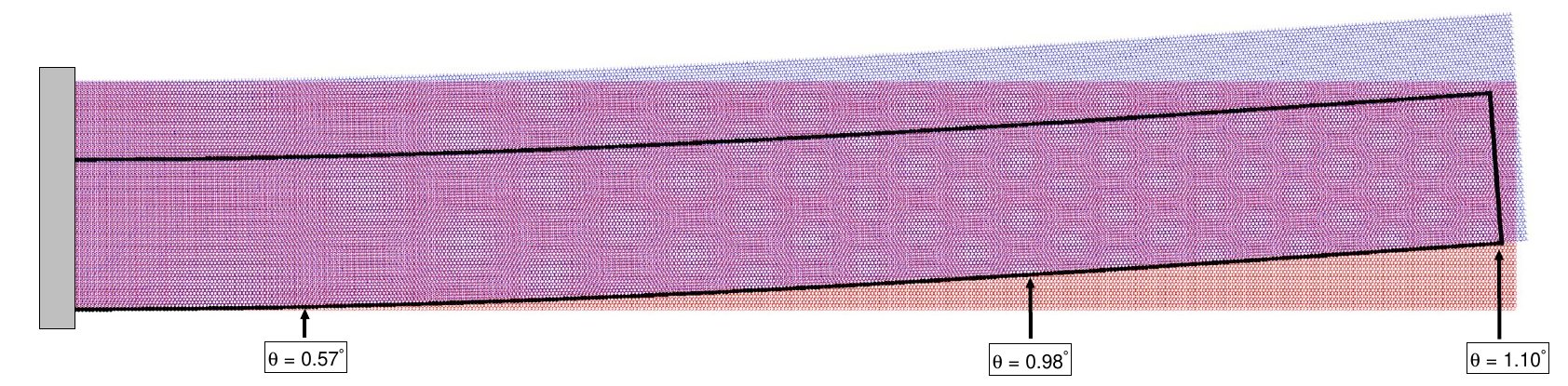}}
\caption{Representation of a bent nanoribbon placed on the top of a graphene substrate (which here is a non-bent ribbon for numerical ease, but in the experiment is a wider graphene layer). The tight binding calculations described in the main text are performed in the entire system including the monolayer fringes (this gives the low-energy DOS shown in blue in Fig. \ref{fig: Figure4} in the main text), then the DOS is projected over the area within the black lines (black curve in Fig. \ref{fig: Figure4}) and the further projected within the region with angles (0.98$^{\circ}$$-$1.10$^{\circ}$) (red curve) or (0$^{\circ}$$-$0.57$^{\circ}$) (orange curve). The positions of these angles are indicated in the figure. In the numerical calculations, the ribbon length is $L \sim 800$ nm and width $W = 80$ nm. The maximum twist angle in the scaled lattice (see main text) is set to $\theta_{max} = 1.1 ^\circ$.}
\label{fig:FigureS1}
\end{figure*}

Figure~\ref{fig:FigureS1} represents a bent nanoribbon analogous to the system used in the tight binding calculations, although smaller for clarity. Within the scaling approximation, the bending angle goes from $0^\circ$  to $1.1^\circ$ as shown in the figure. As seen in the figure, the nanoribbon, bent on top another nanoribbon, leaves some monolayer fringes. 

\section{Scaled tight binding hamiltonian} \label{App: Scaling}

To calculate the electronic spectrum, we model the system with the Lin-Tománek tight binding Hamiltonian \cite{lin2018minimum}:

    \begin{equation}
        H_0 = -\sum_{i\neq j,m}\gamma_{ij}^{mm}(c^{\dagger}_{i,m}c_{j,m}+h.c.) 
        -\sum_{i, j,m}\gamma_{ij}^{m,m+1}(c^{\dagger}_{i,m}c_{j,m+1}+h.c.),
    \end{equation}
where $i,j$ run over the lattice sites and $m$ is the layer index. The Hamiltonian includes intralayer hopping to nearest-neighbours only, accounting for lattice distortions $\gamma_{ij}^{mm}=t_{\parallel}e^{-(r-a_{cc})/\lambda_\parallel}$, where $a_{cc}$ is the nearest neighbour distance. The interlayer hopping decays exponentially away from the vertical direction, $\gamma_{ij}^{m,m+1}=t_{\perp}e^{-(\sqrt{r^2+d_0^2}-d_0)/\lambda_\perp}\frac{d_0^2}{r^2+d_0^2}$, where $d_0=0.335$ nm is the interlayer distance, $t_{\parallel}=3.09$ eV and $t_{\perp}=0.39$ eV are the intralayer and interlayer hopping amplitudes, $\lambda_{\parallel}=0.042$ nm and $\lambda_\perp=0.027$ nm are cutoffs for the intralayer and interlayer hoppings \cite{pereira2009tight,lin2018minimum}. 

\begin{figure*}
    \centering
    {\includegraphics[width=14 cm]{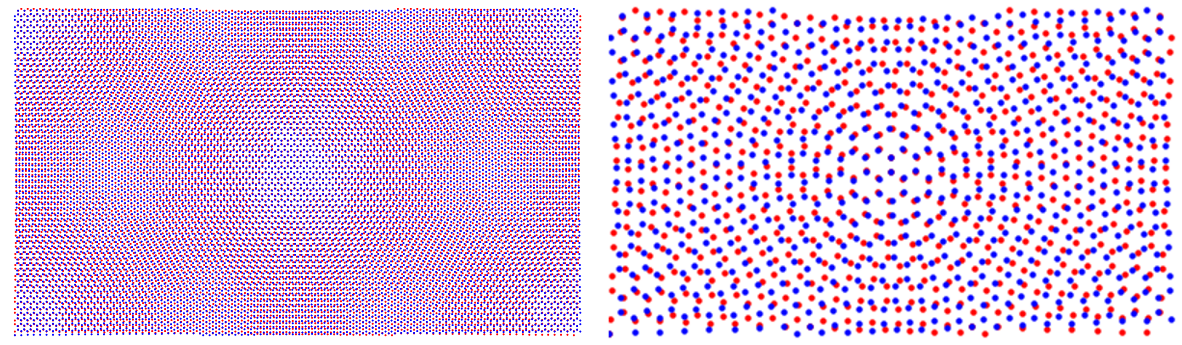}}
    \caption{(Left) Twisted bilayer graphene nanotube unit cell with twist angle 1.12$^{\circ}$. (Right) Scaled unit cell with twist angle 4.41$^{\circ}$.}
    \label{fig: FigureS2}
\end{figure*}

The parameters in the tight binding model are scaled, so that the central bands of TBG with twist angle $\theta$ are approximated by the central bands of an equivalent lattice with twist angle $\lambda \theta$, with $\lambda >1$~\cite{gonzalezarraga2017electrically,vahedi2021magnetism,sainzcruz2021high}. This scaling approximation is based on the fact that the Dirac equation that governs each layer can be described in different ways, and one of them is a honeycomb lattice of super-atoms, each representing a small cluster of atoms of the original lattice. In this scaled system, the lattice constant is magnified and the intralayer hopping reduced. In the context of twisted bilayer graphene, the possibility of scaling manifests also in the continuum model, which shows that the bands depend, to first order, on a dimensionless parameter~\cite{bistritzer2011moire},
\begin{equation}
\alpha = \frac{at_{\perp}}{2\hbar v_F \sin ( \theta / 2 )} \propto \frac{t_{\perp}}{t_0\theta}\, .
\label{eq:s4}
\end{equation}
where $a$ is the lattice constant and $v_F$ is the Fermi velocity. This parameter can be understood as a comparison between the time a carrier needs to traverse a unit cell within a layer, and the average time between interlayer tunneling events.
Thus, a small angle $\theta$ can be simulated with a larger one $\theta^{\prime}$ by doing the following transformations: $t_0\rightarrow\frac{1}{\lambda} t_0$, $a\rightarrow\lambda a$, $d\rightarrow\lambda d$, with $\lambda=\sin(\frac{\theta^{\prime}}{2})/\sin(\frac{\theta}{2})$ \cite{gonzalezarraga2017electrically,vahedi2021magnetism,sainzcruz2021high}. Note that the interlayer distance is scaled to keep the interlayer hopping unchanged after scaling.\ This approximation reproduces well the low-energy band structure, as shown in Fig.~\ref{fig:ScalingBandsComp}, which compares the band structure of twisted bilayer graphene nanotubes~\footnote{In this particular system, imposing periodic boundary conditions in the horizontal direction to close the tube quantizes the corresponding momentum and leads to a folding of the Brillouin zone, which results in 4 flat bands instead of 2 in Figure \ref{fig:ScalingBandsComp}.} with and without scaling, the unit cells are shown in Figure~\ref{fig: FigureS2}. As seen in Figure~\ref{fig:ScalingBandsComp}, the scaled band structure preserves all the qualitative features of the original band structure, except that it suffers a rigid blue-shift. 

\section{Stacking dependence} \label{App: Stack}
\begin{figure*}[htp]
\centering
{\includegraphics[scale = 0.45]{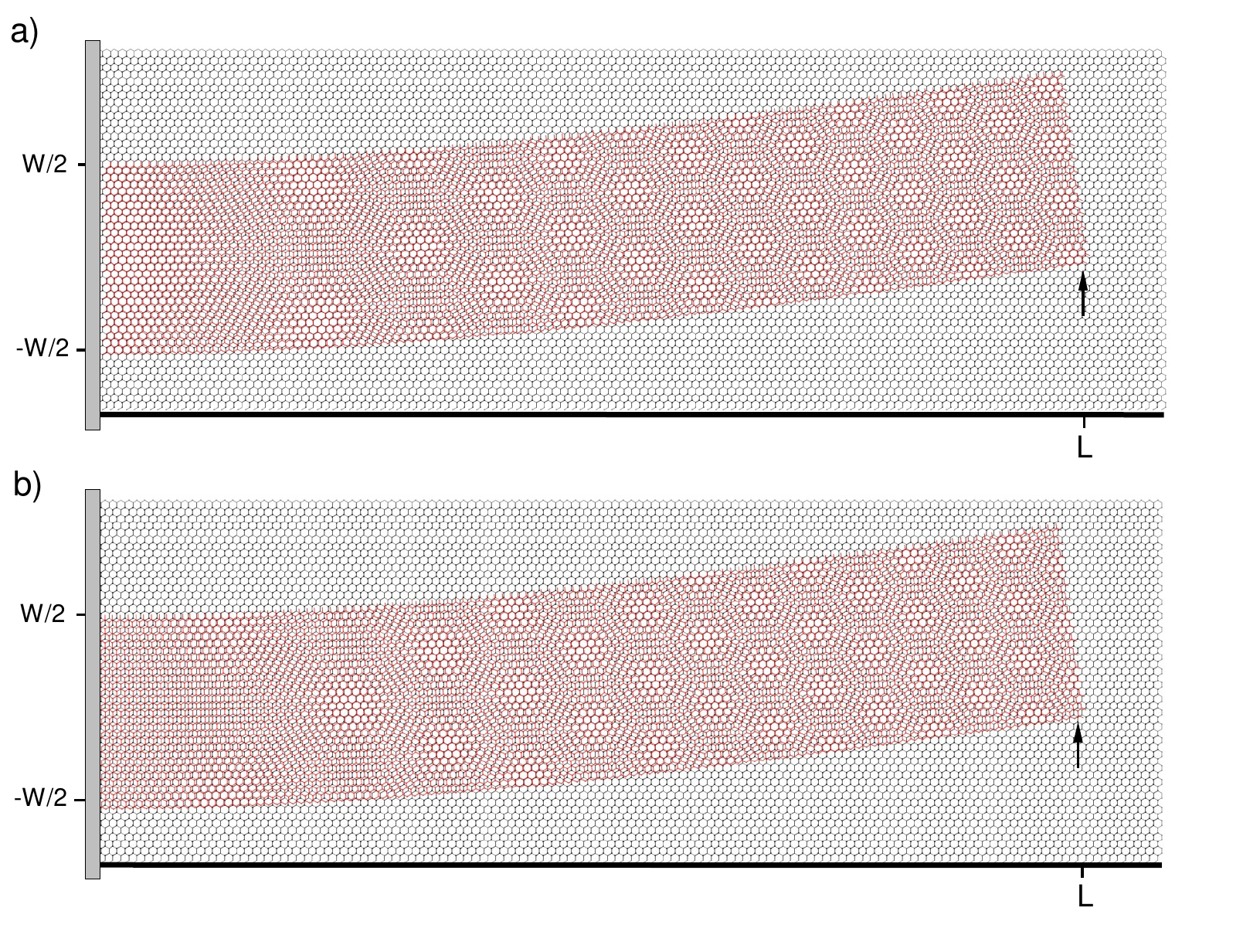}}
\caption{Moir\'e patterns of a bent nanoribbon with length $L$ and width $W$ placed on a graphene substrate. In each panel we show the lattice structure for a) AA and b) AB stack at the $x=0$ origin. The maximum twist angle is set to $8^\circ$. The black arrows indicate the positions of the applied forces as described in the main text.}
\label{fig:SMStack}
\end{figure*}

\begin{figure*}
    \centering
    {\includegraphics[width=13cm]{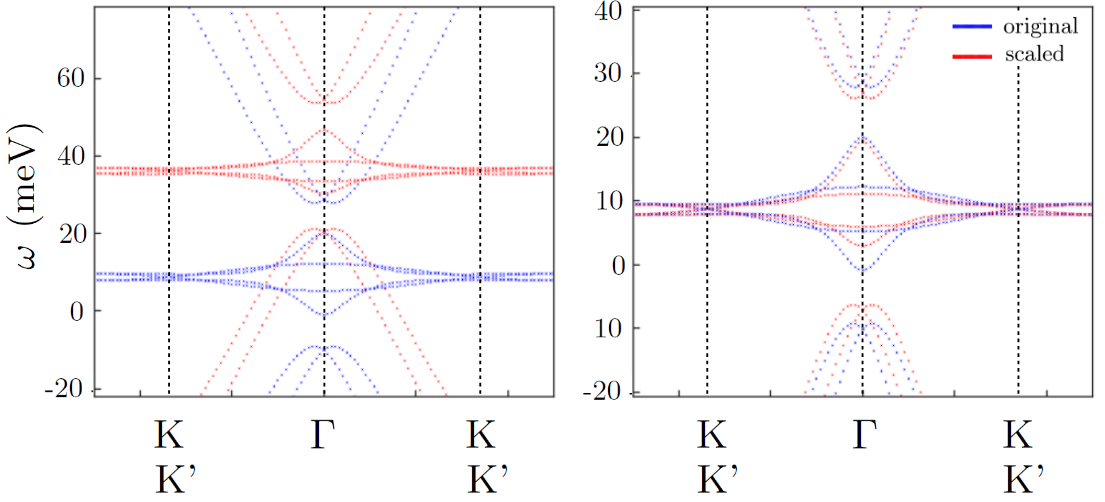}}
    \caption{(Left) Twisted bilayer graphene nanotube bands for a twist angle of 1.12$^{\circ}$, calculated with the lattice in Fig. \ref{fig: FigureS2}(R), i.e. with a scaling approximation (red) and with the lattice in Figure \ref{fig: FigureS2}(L), i.e. with no scaling (blue). (Right) Comparison of bands after equating the Dirac point energies.}
    \label{fig:ScalingBandsComp}
\end{figure*}

In Fig.~\ref{fig:SMStack} we display the lattice structure of a bent nanoribbon with two different stackings. The maximum twist angle is set to $8^\circ$. Figure~\ref{fig:SMStack}a) is for an AA and Fig.~\ref{fig:SMStack}b) for AB stack at the clamped $x=0$ origin.  

\section{Other charge maps} \label{App: ChargeMaps}
\begin{figure*}[h]
\centering
\includegraphics[scale = 0.73]{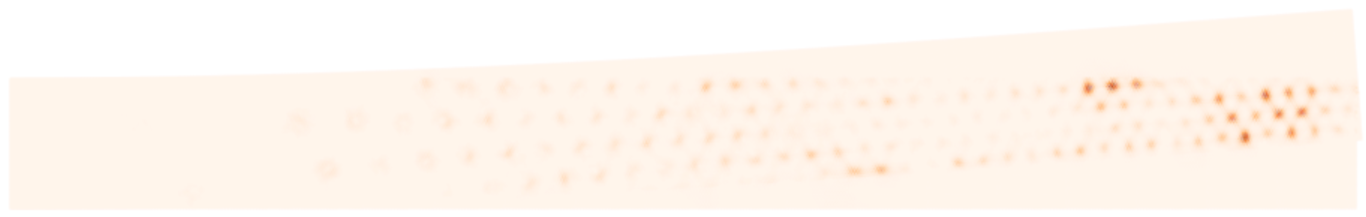}
\caption{Charge map of a state with $E=35.59$ meV.}
\label{fig: FigureS5}
\end{figure*}

\begin{figure*}[h]
\centering
\includegraphics[scale = 0.73]{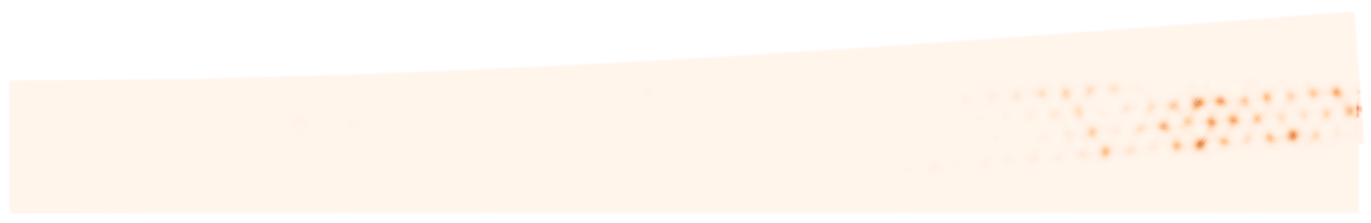}
\caption{Charge map of a state with $E=36.38$ meV.}
\label{fig: FigureS6}
\end{figure*}

\begin{figure*}[h]
\centering
\includegraphics[scale = 0.73]{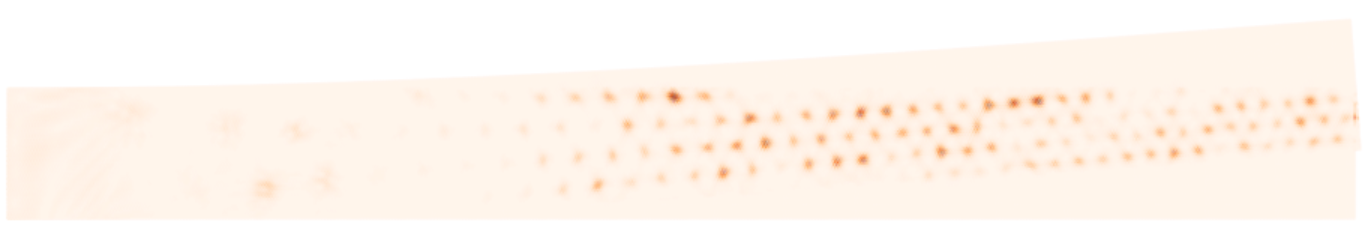}
\caption{Charge map of a state with $E=39.32$ meV.}
\label{fig: FigureS7}
\end{figure*}

\begin{figure*}[h]
\centering
\includegraphics[scale = 0.73]{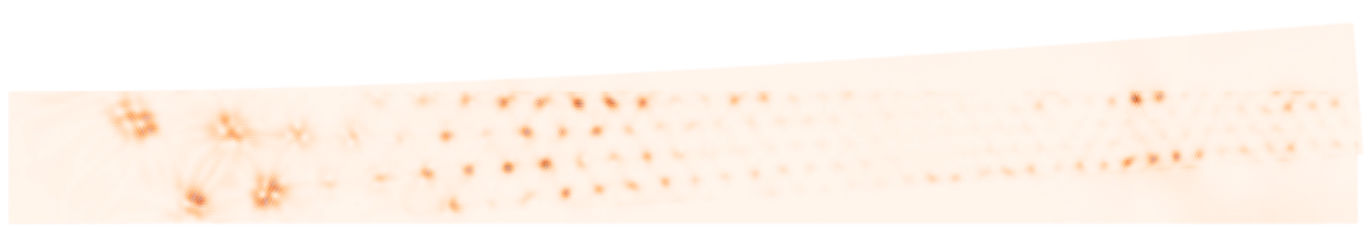}
\caption{Charge map of a state with $E=46.95$ meV.}
\label{fig: FigureS8}
\end{figure*}

\begin{figure*}[h]
\centering
\includegraphics[scale = 0.73]{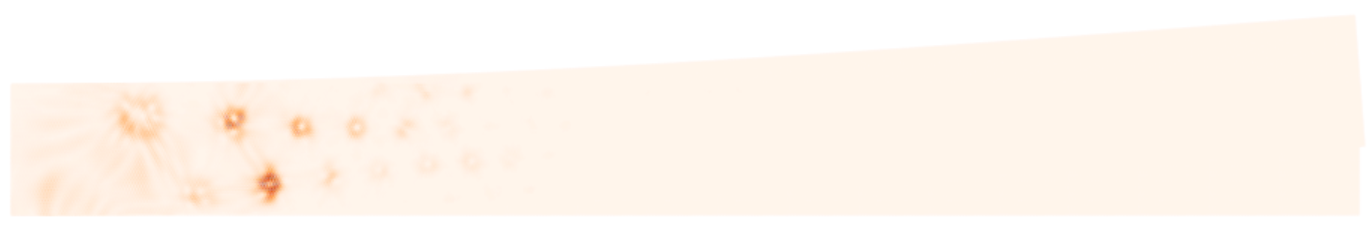}
\caption{Charge map of a state with $E=53.7$ meV.}
\label{fig: FigureS9}
\end{figure*}

\begin{figure*}[h]
\centering
\includegraphics[scale = 0.73]{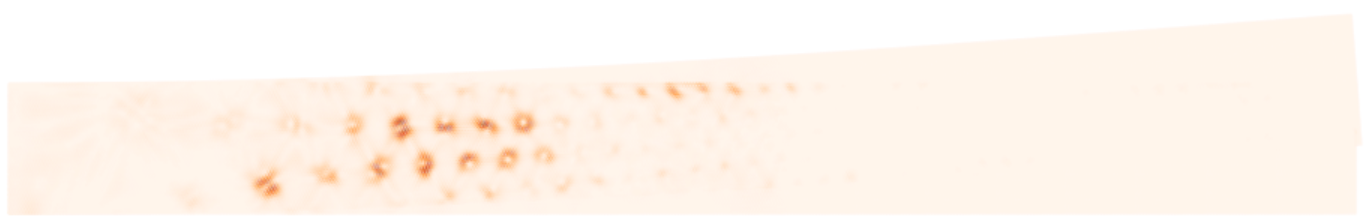}
\caption{Charge map of a state with $E=57.27$ meV.}
\label{fig: FigureS10}
\end{figure*}

Figures~\ref{fig: FigureS5} to~\ref{fig: FigureS10} show charge maps for states with the indicated energies, cf. Fig.~\ref{fig: Figure4}. Many states show charge localization at the AA stacking regions, suggesting that a bent nanoribbon is ideal to explore correlated effects. Away from the DOS peak (which is at $E\approx40$ meV), there are states with annular charge localization around the AA stacking regions.

\end{document}